# Exploring the Links between Personality Traits and Susceptibility to Disinformation

Dipto Barman
School of Computer Science and Statistics
Trinity College Dublin
Dublin, Ireland
barmand@tcd.ie

Owen Conlan
School of Computer Science and Statistics
Trinity College Dublin
Dublin, Ireland
Owen.Conlan@scss.tcd.ie

## ABSTRACT

The growth of online Digital/social media has allowed a variety of ideas and opinions to coexist. Social Media has appealed users due to the ease of fast dissemination of information at low cost and easy access. However, due to the growth in affordance of Digital platforms, users have become prone to consume disinformation, misinformation, propaganda, and conspiracy theories. In this paper, we wish to explore the links between the personality traits given by the Big Five Inventory and their susceptibility to disinformation. More specifically, this study is attributed to capture the short-term as well as the long-term effects of disinformation and its effects on the five personality traits. Further, we expect to observe that different personalities traits have different shifts in opinion and different increase or decrease of uncertainty on an issue after consuming the disinformation. Based on the findings of this study, we would like to propose a personalized narrative-based change in behavior for different personality traits.

## 1   INTRODUCTION

The boom of Digital/social media and the decline of traditional news outlets has led more people getting their information online. A survey conducted by the Pew Research center in 2016 indicated that 62% of US adults get their daily news from social media [6], increasing from 49% in 2012. These social media platforms provide users to express, share, and interact with content and other users with ease. These changes have made the consumption of information more affordable, but it has also enabled bad actors to push disinformation in the digital space. This increase of affordance of social media, users have become prone to consume disinformation, misinformation, propaganda, and conspiracy theories. Termed as "Infodemic" by the World Health Organization (WHO) [20], it has left many individuals baffled about what exactly is the truth. A perfect example of verified and unverified claims are conflicting reports, hoaxes, and conspiracy theories that surround the Covid-19 pandemic [12]. As defined in [9], misinformation is "*false or misleading information*" and disinformation is "*False information that is purposely spread to deceive people*". Consequently, these information disorders have strained researchers, policymakers, scholars, and others to provide countermeasures.

## 2   RELATED WORK

Serval studies have been conducted around disinformation and its mitigation. In literature, there are two approaches to counter disinformation, namely proactive and reactive. Reactive approaches are related with the efficacy of debunking and fact-checking. As the information spreads in the digital space, these reactive approaches come into play. However, certain concerns have risen with respect to reactive approaches. A plethora of research suggest that the mere repetition of information can increase its perceived truthfulness and allow rejection of new information that sheds an honest light on the topic [5]. For example, the repeated surveys done by the *Washington Post* found that even after correcting misleading and false statements of Donald Trump at multiple instances, republicans considered President Trump to be honest. This highly questions the efficacy of fact-checking, as even with fact-checkers in place, these corrections fail to alter people's opinion about misleading information [10]. Another problem with fact-checking is just the pace of checking. Due to increase of affordance of digital media, the spread of disinformation has outpaced fact-checkers and thus poses a different solution that needs to be more proactive in nature, rather than reactive.

Scholars have proposed several proactive intrusions to counter disinformation. The idea behind proactive approaches is to reduce the believability and sharing of disinformation in the first place. Pre-bunking is process of inoculation based on the inoculation theory [13], where a threat message combined with preemptive refutation is provided to individuals to develop a resistance against the persuasion of disinformation. Several body of inoculation research has shown its effectiveness against disinformation (for review see [10]). In [21], the authors proposed a study to scrutinize the effects of online astroturfing for digital enabled foreign propaganda and found that inoculation treatments, particularly refutation-same treatment were able to inoculate the study-group against a foreign propaganda. They also found that after 2 weeks, refutation-same treatment was more potent in reducing opinion changes. In [11], the authors inoculated the study group and were able to confer resistance against misinformation about climate change. One of the most critically acclaimed examples of inoculation is an interactive social media game called *Bad News* [16], which actively inform the participants about the techniques that are used to spread disinformation.

However, the previous research focused on "one size fits all" solution for disinformation. As stated in [21], "*disinformation is a form of persuasion*". Research in the domain of persuasive technologies [1, 7, 14] has indicated that personalized approaches to individuals provide a better response for persuasion rather than "one size fits all" type solution. As stated in [18], personal efficacy is one of the reasons why an individual reacts to a fake news. However, to the best of our knowledge no principled study has been conducted on susceptibility of disinformation on personality traits. Therefore, to give a comprehensive understanding, we investigate the following three research questions:

- **RQ1**: Which users are more likely to get influenced by disinformation?
- **RQ2**: What are the personality traits of the users that are more likely get influenced by disinformation, and do they have a clear difference?
- **RQ3**: Do the effects of disinformation persist after two weeks, given the choice of investigating the disinformation itself?

By investigating **RQ1**, we hope to identify users who are more likely to get influenced by disinformation, which can be treated as a representative user group to characterize personality traits. By answering **RQ2**, we expect to provide guidance on assessing whether different personality traits are affected differently with respect to the same attack message. By analyzing **RQ3**, we hope to study the effects of disinformation over a period of two weeks on different personality traits.

**Personality Traits.** Personality traits are a key distinct feature that defines behavior, emotions, and interactions of an individual and is related to preference and interests. The big five factor model [8] (also known with acronym "OCEAN") has been utilized in persuasive technologies and in information system research. The description of each personality trait can be found in table 1. It has been widely used to study the relationship between persuasive techniques and a promoting healthy lifestyle decision such as sustainable transportation [1], health promoting Application [7], and health promoting games [14]. Personality traits have also been used to study the ability to discern news. In [19], Wolverton and Stevens reported that extraversion and openness personality traits were negatively corelated to fake news. However, their results should be interpreted cautiously because their analysis was based on a dichotomized response to single personality factor items and fake news identification was measured with users selecting which of the nine headlines, they believed were fake. Most discernment of fake news studies are based on rating the accuracy of the headline on a scale to measure belief in a more sensitive manner than dichotomous judgements [2, 3, 15]. However, to come to conclusion that the ability to judge a news article just by reading relates to susceptibility to disinformation seems a bit far-fetched. As stated in [18], individuals react to fake news based on the relevance of the issue. And it is possible that the news articles used in [2, 3, 15, 19] might not be of relevance to participants, and the ability to discern news might just be based on chance. Hence, the results from [2, 3, 15, 19] should be cautiously interpreted. Thus, we propose to evaluate explicit personality factors given by the OCEAN model and the change in opinion and opinion certainty based on an attack message.

**Table 1: Personality Traits [8]**

| Personality Traits | Characteristics |
| --- | --- |
| Openness | Open to trying new things, Creative, likes to think about abstract concepts. |
| Conscientiousness | Prominent level of thoughtfulness, good impulse control and goal oriented. |
| Extraversion | Talkative, assertiveness, and considerable amount of emotional expressiveness. |
| Agreeableness | Prosocial behavior, cooperative, has great deal of interest in other people. |
| Neuroticism | High mood swings, irritability, and sadness. |

## 3 METHODOLOGY

We hope to conduct our experiment in two sub experiments. Participants are to be recruited through a commercial online access panel (*Prolific Academic*) in the coming months. Using G* power [4], a sample of at least 345 participants would be required to provide a power of 0.85 (with two-tailed $\alpha = 0.05$). We opt to conduct our study in the global context, primarily for reasons of demographic relevance. For our attack message, we have selected the conspiracy theory "*Covid-19 is a Man-Made Virus made in a Wuhan Lab by Chinese Scientists*" [22]. We have selected this attack message as it is a global phenomenon, and on the assumption that every participant has experience Covid-19 in one form or another. We made this choice based on the findings reported in [18], as Covid-19 has had a significant impact in our daily life. As this conspiracy theory is picking up pace again, we believe that this attack message would provide a significant response to our research questions.

**Experiment.** The experiment is designed into two sub-experiments to study the effects of short-term and long-term effects of disinformation on different personality traits. In the first phase of sub-experiment 1, the participant personality trait is to be recorded using the Big Five Inventory-2-S (BFI-2-S) [17]. The BFI-2-S contains 30 statements in which participants would rate their agreement on a scale of 1 (*Strongly Disagree*) to 5 (*Strongly Agree*). The five personality factors: Openness, Conscientiousness, Extraversion, Agreeableness, and Neuroticism are coded in six statements, half being reverse coded. We adopted the method employed in [21], and before the commencement of the second phase, participants would be asked to indicate their opinion on the question "*Do you believe that Covid-19 was made in a Lab in China?*" and their belief is recorded on a 5 point-Likert scale (1 "*I Strongly believe that Covid-19 was made in a lab in China*" to 5 "*I strongly disbelieve that Covid-19 was made in a lab in China*"). Similarly, participants would also be asked to rate the opinion certainty using a 5-point Likert scale (1 "*Not certain at all*" to "*Extremely Certain*"). In the second phase, the attack message "*Covid-19 is a Man-Made Virus made in a Wuhan Lab by Chinese Scientists*" news article [23] would be shown to participants. The attack message is subject to change, if a more convincing article with rhetoric's is found. In the final phase, the same question as in second phase will be asked to record the opinion and opinion certainty. Similar to sub-experiment 1, sub-experiment 2 follows the same roadmap till the commencement of phase 3. After the display of the attack message, the participants would be asked to investigate the attack message as per their convivence using any source if they wish in a two weeks' time. After two weeks, the participants would then be asked same question as in second phase to record their opinion and opinion certainty after going through a manipulation check which indicates if the participant remembers what the attack message was about and if he has checked our sources to validate the claims of the attack message. Finally, in both the sub-experiments, a post demographic questionnaire would be asked for validating the sample population. The flow diagram for the experiment is illustrated in Figure 1.

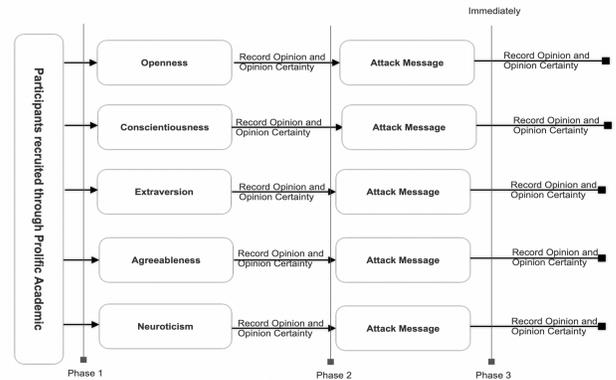

**Figure 1: Experiment Flow Diagram for the Study.**

## 4 DATA ANALYSIS

After the conclusion of the two sub-experiments, subtracting participants' post-stimulus from their pre-stimulus answers would result in two scores which would reflect the changes in opinion and opinion certainty. Positive values of the opinion-change would indicate that participants did believe that "that Covid-19 was made in a lab in China" after seeing the attack message. Positive values of the opinion-certainty-change measure would indicate that participants grew more uncertain compared to their initial certainty assessment.

First, we would need establish the effect size of independent variable (personality traits) over the two-dependent variable "opinion change" and "opinion-certainty-change". We hope to see at least a medium size effect with f>0.25 using IBM SPSS statistics v27 tool between the independent variables and the two dependent variables. From sub-experiment 1, using the one-way ANOVA test, we hope to see a statistical difference between the personality trait groups (independent variable) with respect to opinion change, opinion-certainty-change (dependent variables), answering RQ1. Using linear regression, we hope to find some significant correlation between different personality traits and difference in opinion, opinion-certainty change, which would consequently answer RQ2, stating that the same message effects differently personality traits differently. From sub-experiment 2, using linear regression, we hope to see the long-term effects of attack message vary differently with different personality trait answering our RQ3. It is also possible that our results from sub-experiment 1 might differ from sub-experiment 2 given that the individuals were given time to assess the attack message in a two weeks' time.

## 5 CONCLUSION

It is clear through our literature survey that disinformation is a highly complex area that has no direct solutions. The study mentioned above would help elucidate the relationship between personality traits, change in opinion, and change in opinion certainty due to disinformation on social media. We hypotheses to see that certain personality traits may be highly correlated to change in opinion, resulting in higher susceptibility to disinformation. Using these results from this study as groundwork, we hypothesis to study and develop personalized narrative-based intervention techniques for different individuals on social media. It would lay the foundation for future research in field of countering susceptibility to disinformation and adaptive personalized inoculation techniques.

## ACKNOWLEDGMENTS

This work was conducted with the financial support of the Science Foundation Ireland Centre for Research Training in Digitally-Enhanced Reality (d-real) under Grant No. 18/CRT/6224 and at the ADAPT SFI Research Centre at Trinity College Dublin. ADAPT, the SFI Research Centre for AI-Driven Digital Content Technology, is funded by Science Foundation Ireland through the SFI Research Centres Programme and is co-funded under the European Regional Development Fund (ERDF) through Grant Agreement No. 13/RC/2106.